\documentclass[superscriptaddress,aps,pra,twocolumn,showpacs]{revtex4-2}

\usepackage[latin1]{inputenc}
\usepackage{graphicx,amsmath,amsfonts,amssymb}
\usepackage{color}
\usepackage[colorlinks=true]{hyperref}
\usepackage{graphicx}
\usepackage{epstopdf}
\usepackage{bbm}

\newcommand{\ket}[1]{\mbox{$ | #1 \rangle $}}
\newcommand{\bra}[1]{\mbox{$ \langle #1 | $}}
\newcommand{\be}{\begin{equation}}
\newcommand{\ee}{\end{equation}}
\newcommand{\beq}{\begin{eqnarray}}
\newcommand{\eeq}{\end{eqnarray}}
\newcommand{\expval}[1]{\mbox{$\langle #1 \rangle$}}

\newcommand{\orcid}[1]{\href{https://orcid.org/#1}{\includegraphics[width=7pt]{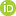}}}

\begin{document}

\title{Realism-based nonlocality: Invariance under local unitary operations and asymptotic decay for thermal correlated states}

\author{V. S. Gomes\orcid{0000-0001-6768-7837}}
\affiliation{Department of Teleinformatics, Federal University of Cear\'{a}, Fortaleza, CE, Brazil}

\author{P. R. Dieguez\orcid{0000-0002-8286-2645}}
\affiliation{Center for Natural and Human Sciences, Federal University of ABC, Avenida dos Estados $5001, 09210-580$, Santo Andr\'{e}, SP, Brazil}

\author{H. M. Vasconcelos\orcid{0000-0001-7129-1026}}
\affiliation{Department of Teleinformatics, Federal University of Cear\'{a}, Fortaleza, CE, Brazil}

\begin{abstract} 
The realism-based nonlocality (RBN) is a recently introduced measure that differs from the well-known Bell's nonlocality. For bipartite states, the RBN concerns how much an element of reality associated with a given observable is affected upon local measurements on a subsystem.  Here, we present an analytical proof for the unitary invariance of the RBN and that it presents a monotonous behavior upon the action of unital and non-unital local quantum noise. We illustrate our results by employing the two-qubits Werner state and thermal quantum correlated states. We show how the RBN is limited by the initial equilibrium temperature and, especially, that it decays asymptotically with it. These results also corroborate the hierarchy relationship between the quantifiers of RBN and global quantum discord, showing that RBN can capture undetectable nonlocal aspects even for non-discordant states. Finally, we argue how our results can be employed to use the RBN as a security tool in quantum communication tasks.

\end{abstract}

% \PACS{PACS 03.67.Mn \and PACS 03.65.Ud \and PACS 03.67.-a}
\maketitle

%=================================================
\section{Introduction}
\label{intro}

Under the premise of locality, Einstein, Podolsky, and Rosen (EPR) denied the completeness of quantum theory by arguing that the realism of incompatible observables can be simultaneously determined in scenarios involving entangled states~\cite{EPR}.  The EPR's criterion~\cite{EPR} for physical realism makes direct reference to eigenstates of the observables being measured and it is related to certainly predict,  without disturbing,  the value of some physical property.  However, as pointed out by Bell \cite{Bell,BrunnerWehner}, and confirmed by loopholes-free experiments \cite{hensen15,giustina15,shalm15}, the correlations observed in isolated microscopic systems cannot be reproduced by any theory supplemented with hidden local causal variables. Today, we know that these are effects that manifest themselves through the widely known quantum correlations \cite{CostaAngelo,WisemanDoherty}, which play a key role in modern quantum information theory and are fundamental in many practical applications in quantum computing.

Inspired by EPR's \cite{EPR} definition of elements of reality, Bilobran and Angelo have proposed a measure capable of quantifying the degree of realism of an observable for a given preparation which is based on a single measurement premise~\cite{Bil}. This measure has proven to be relevant in scenarios involving weak measurements~\cite{Monitoring}, quantum walks \cite{NBRQuantumWalking}, resource theories \cite{costa20}, matter-wave interferometry~\cite{Lustosa20}, and more recently with an experimental investigation in the context of Wheeler's delayed-choice experiments and wave and particle duality~\cite{Dieguez21}. Furthermore, it was possible to establish a concept of nonlocality that is different from Bell's nonlocality, which is known as realism-based nonlocality (RBN) \cite{NBR}. Unlike Bell's nonlocality, which is verified through inequality violations based on the causal locality hypothesis, RBN occurs when there are changes in the degree of realism due to local operations occurring in a remote location. The RBN has also shown resilience \cite{NBRResilient} in situations involving local and bilocal weak measurements through a process called monitoring \cite{Monitoring,dieguez2018}, being the most robust form of quantumness considering a wide known class of quantum correlations. In addition, it was also investigated in tripartite states \cite{NBRTripartite}. 

Quantifiers of quantum correlations have some basic requirements~\cite{PopescuRohrlich, GuifreVidalTarrach}, as for example, to quantify entanglement, it is expected that this measure vanishes for separable states, be invariant under local unitary operations \cite{BennettWootters,MonrasIlluminati}, and be monotone under local operations~\cite{VedralPlenio,GuifreVidal,Horodecki}. In the case of quantifiers of Bell nonlocality \cite{Bell,BrunnerWehner}, there are works \cite{BritoAmaralChaves, JulioVicente, GallegoAolita} that established some criteria to be verified.  Moreover, quantum discord \cite{OllivierZurek,Henderson01,RulliSarandy} has shown to be a type of quantum correlation that can occur even for non-entangled states. In this context, together with EPR steering, all these quantum correlations are connected through a hierarchy relation of quantumness measures constructed in such a way that the discordant states form a strict superset of all other measures \cite{WisemanDoherty,NBR,NBRResilient}, with such hierarchy being maintained even under the effect of quantum noise \cite{CostaAngelo}.

In this work, we first employ the RBN and present analytical proof that guarantees the invariance under local unitary operations. We investigate the potential use of that measure for quantum information and thermodynamic protocols since the quantum correlations presented in thermal states are usually associated with some energetic resource~\cite{Sapienza19,X_State} and can be useful to quantum thermal communication tasks~\cite{NewtonI,NewtonII}.  Following the result obtained in Ref.~\cite{NBRResilient} that proved the monotonous relation under the unital monitoring map~\cite{Monitoring},  we explore here the monotonous behavior of RBN under both unital and non-unital quantum noise channels that act locally on one of the subsystems.  We verify this for the well-known bipartite Werner state to illustrate the result of the applied noisy channels and the hierarchy relation among RBN, quantum discord, and entanglement. Finally, we investigate how RBN behaves concerning limitations of the local temperature of two uncorrelated Gibbs states which is submitted to an optimally unitary operation to maximize the quantum correlation between them~\cite{X_State}, the behavior under noisy channels, and hierarchical relations among quantum correlations. 
This work is structured as follows. In Section II we briefly review the measures relevant to our work and stress how the notion of elements of reality are related to this particular measure of nonlocality. In Section III, we present the unitary invariance of RBN and its monotonicity under local operations. In section IV, we show that RBN can guarantee the security of quantum correlated channels with noise, thus showing that it can be used for secure communication. In Section V we explore the developed result in the context of quantum correlated thermal states. Finally, in Section VI we draw our conclusions.

\section{Realism-based nonlocality}
 
The RBN measure is based on the so-called irrealism which establishes the degree of realism for an observable given a certain preparation~\cite{Bil}. The irrealism measure was constructed by assuming a single criterion that projective unread measurements establish the reality of that measured observable. This is formalized by considering a bipartite state $\rho$ which acts in $\mathcal{H}=\mathcal{H}_{\mathcal{A}}\otimes \mathcal{H}_{\mathcal{B}}$ that is non-selective measured by an observable $ A $ acting on $ \mathcal{H}_{\mathcal{A}} $, resulting in the average post-measured state
\begin{equation}
\Phi_{A}(\rho) := \sum_a (A_{a} \otimes \mathbbm{1}^{\mathcal{B}}) \rho (A_{a} \otimes \mathbbm{1}^{\mathcal{B}}) = \sum_a p_{a} A_{a} \otimes \rho_{\mathcal{B}|a}
\end{equation}
where $\rho_{\mathcal{B}|a} = \mathrm{Tr}_{\mathcal{A}} (A_{a} \otimes \mathbbm{1}^{\mathcal{B}}\rho)/p_{a}$ and $p_{a} = \mathrm{Tr} (A_{a} \otimes \mathbbm{1}^{\mathcal{B}})$. According to the reality criteria proposed in Ref.~\cite{Bil}, the map $\Phi_{A}(\rho)$ can be understood as a state of reality defined for $A$.
This criterion led the authors to define the irreality of observables, a quantifier for the degree of irrealism of an observable
\begin{equation}
\Im(A|\rho):= S(\Phi_{A}(\rho)) - S(\rho),
\end{equation}
where $S$ stands for the von Neumann entropy, $A=\sum_{a}aA_{a}$ a observable with projectors $A_{a} = |a\rangle \langle a|$ defined in terms of the eigenvectors of $A$. This measure is non-negative, $\Im(A|\rho)$ $\geqslant$ $0$,
%since unread projective measurements never decreases entropy,
being null only for states that are already a state of reality for that observable, this is, $\Phi_{A}(\rho) = \rho$.
Based on the premise above, in Ref.~\cite{Bil} it was also defined the contextual nonlocality
\begin{equation}
\eta_{AB}(\rho):= \Im(A|\rho) - \Im(A|\Phi_{B}(\rho)),
\label{eq3}
\end{equation}
a measure capable of quantifying, given the context $(A,B,\rho)$, how irreality of $A$ changes since measurements of $B$ are conducted in a remote location.
Among some mathematical properties of $ \eta_{AB}(\rho)$ are the invariance under index exchange $A\rightleftarrows B$, which is noticed when we write the Eq. \ref{eq3} in terms of the von Neumann entropy
\begin{equation}
\eta_{AB}(\rho) = S(\Phi_{A}(\rho)) + S(\Phi_{B}(\rho)) - S(\Phi_{AB}(\rho)) -S(\rho),
\label{eq4}
\end{equation}
and the non-negativity $\eta_{AB}(\rho) \geqslant 0$, with equality holding for uncorrelated states $\rho =\rho_{A}\otimes\rho_{B}$, 
and for states that satisfy the reality criterion, that is, $\Phi_{A}(\rho) = \rho$, $\Phi_{B}(\rho) = \rho$ or $\Phi_{AB}(\rho) = \rho$.
 
 In regard to all the possible contexts $(A,B,\rho)$, it was introduced in Ref.~\cite{NBR} the concept of realism-based nonlocality (RBN)
\begin{equation}
N_{\mathrm{rb}}(\rho) := \max_{A,B}\eta_{AB}(\rho),
\label{eq5}
\end{equation}
a quantifier capable of capturing the entire RBN contained in a quantum state. This quantifier, clearly, is independent of any context $(A,B,\rho)$ considered, depending only on the quantum state. In addition, it is non-negative, $N_{\mathrm{rb}}\geqslant 0$, since $\eta_{AB}(\rho) \geqslant 0$, and is null only for uncorrelated states $\rho =\rho_{A}\otimes\rho_{B}$ and for states of reality $\Phi_{R}(\rho)=\rho$ with $R =A,B,AB$. 
An interesting aspect of $N_{\mathrm{rb}}$ refers to what is known as a nonlocality anomaly, a problem associated with the Bell \cite{AcinEtAl, AnomalyNonlocality, VolumeViolacao} nonlocality quantification framework.
Interestingly, no anomaly manifests itself in the RBN quantification scenario, which tells us that a maximally entangled state, $|\phi\rangle=\sum_{i = 1}^{d}|i\rangle|i\rangle/\sqrt{d}$, is also maximally nonlocal.

\section{Local quantum operations}

 Here, we develop analytically and numerically evidences that the RBN measurement does not increase under local quantum operations, which allows its application in quantum information processing tasks. We begin the analysis by presenting a proof that guarantees the local unitary invariance of RBN. Consider a state $\tilde{\rho} = U\rho U^{\dagger}$ acting on a bipartite space $\mathcal{H}_{\mathcal{A}} \otimes \mathcal{H}_{\mathcal{B}}$, where $U = (U_{\mathcal{A}} \otimes U_{\mathcal{B}})$ 
is the local unitary transformation that acts on the initial state $\rho$ and satisfy the condition $U^{\dagger}U = UU^{\dagger} = \mathbbm{1}$. Considering also the definition of RBN (\ref{eq5}) for that state $\tilde{\rho}$, we have that
\begin{equation}
N_{\mathrm{rb}}(\tilde{\rho}) := \max_{A,B} \eta_{AB}(\tilde{\rho}),
\label{eq6}
\end{equation}
with $\eta_{AB}(\tilde{\rho}) = S(\Phi_{A}(\tilde{\rho})) + S(\Phi_{B}(\tilde{\rho})) - S(\Phi_{AB}(\tilde{\rho})) - S(\tilde{\rho})$, where $A = \sum_{a} a A_{a}$ and $B = \sum_{b} b B_{b}$ with projectors $A_{a} = |a \rangle \langle a|$ e $B_{b} = |b \rangle \langle b|$ observables that acts on $\mathcal{H}_{\mathcal{A}}$ and $\mathcal{H}_{\mathcal{B}}$, respectively. Note that
\begin{equation}
\Phi_{A} (\tilde{\rho}) = \sum\limits_{a} (A_{a} \otimes \mathbbm{1}^{\mathcal{B}}) \, \tilde{\rho}\, (A_{a} \otimes \mathbbm{1}^{\mathcal{B}}),
\end{equation}
in an equivalent way as $\Phi_{B}(\tilde{\rho})$ and $\Phi_{AB}(\tilde{\rho})$. Also, we have
\begin{eqnarray}
U^{\dagger} \Phi_{A} (\tilde{\rho}) U &=& U^{\dagger} \left[ \sum\limits_{a} (A_{a} \otimes \mathbbm{1}^{\mathcal{B}}) \, U\rho U^{\dagger}\, (A_{a} \otimes \mathbbm{1}^{\mathcal{B}}) \right] U \nonumber \\ &=& \sum_{a}  (\tilde{A_{a}} \otimes \mathbbm{1}^{\mathcal{B}}) \, \rho \, (\tilde{A_{a}} \otimes \mathbbm{1}^{\mathcal{B}}) \nonumber \\ &=& \Phi_{\tilde{A}} (\rho),
\end{eqnarray}
where $\tilde{A_{a}} = U_{\mathcal{A}}^{\dagger}A_{a}U_{\mathcal{A}}$. Note that, as $U_{\mathcal{A}}$ is a transformation that satisfies the condition $U_{\mathcal{A}}^{\dagger}U_{\mathcal{A}} = U_{\mathcal{A}}U_{\mathcal{A}}^{\dagger} = \mathbbm{1}$, so
\begin{equation}
U^{\dagger} \Phi_{A} (\tilde{\rho}) U = \Phi_{\tilde{A}} (\rho) \ \rightarrow \ \Phi_{A} (\tilde{\rho}) = U \Phi_{\tilde{A}} (\rho) U^{\dagger}. \nonumber
\end{equation}
Similarly, it can be seen that $U^{\dagger} \Phi_{B} (\tilde{\rho}) U  = \Phi_{\tilde{B}} (\rho) \ \rightarrow \ \Phi_{B} (\tilde{\rho}) = U \Phi_{\tilde{B}} (\rho) U^{\dagger}$ e $U^{\dagger} \Phi_{AB} (\tilde{\rho}) U  = \Phi_{\tilde{A}\tilde{B}} (\rho) \ \rightarrow \ \Phi_{AB} (\tilde{\rho}) = U \Phi_{\tilde{A}\tilde{B}} (\rho) U^{\dagger}$. Using the property of invariance of von Neumann entropy under unitary evolution, we find that
\begin{eqnarray}
S(\tilde{\rho}) &=& S(U\rho U^{\dagger}) = S(\rho) \nonumber\\
S(\Phi_{A} (\tilde{\rho})) &=& S(U \Phi_{\tilde{A}} (\rho) U^{\dagger}) = S(\Phi_{\tilde{A}} (\rho))\nonumber\\
S(\Phi_{B} (\tilde{\rho})) &=& S(U \Phi_{\tilde{B}} (\rho) U^{\dagger}) = S(\Phi_{\tilde{B}} (\rho))\nonumber\\
S(\Phi_{AB} (\tilde{\rho})) &=& S(U \Phi_{\tilde{A}\tilde{B}} (\rho) U^{\dagger}) = S(\Phi_{\tilde{A}\tilde{B}} (\rho)).\nonumber
\end{eqnarray}
Now, according to Eq. \ref{eq6}, we have
\begin{eqnarray}
N_{\mathrm{rb}}(\tilde{\rho}) &=& \max_{A,B}\left[ S(\Phi_{A}(\tilde{\rho})) + S(\Phi_{B}(\tilde{\rho})) - S(\Phi_{AB}(\tilde{\rho})) - S(\tilde{\rho}) \right] \nonumber \\ &=& \max_{\tilde{A},\tilde{B}}\left[ S(\Phi_{\tilde{A}}(\rho)) + S(\Phi_{\tilde{B}}(\rho)) - S(\Phi_{\tilde{A}\tilde{B}}(\rho)) - S(\rho) \right] \nonumber \\ &=& N_{\mathrm{rb}}(\rho),
\end{eqnarray}
where the maximization over all observables $\tilde{A}$ and $\tilde{B}$ is equivalent to the same process on $A$ and $B$, because $A = \sum_{a}aA_{a} \ \rightarrow \ U_{\mathcal{A}}^{\dagger}AU_{\mathcal{A}} = \sum_{a}a\tilde{A_{a}} = \tilde{A}$, such that $A$ and its unitary transformation $U_{\mathcal{A}}^{\dagger}AU_{\mathcal{A}} = \tilde{A}$ are equivalent in respect to the maximization process (see \cite{Sakurai} for more details), the same happens for $B$ and $\tilde{B}$. 
Therefore, we see that the condition of invariance under unitary transformations is satisfied by the RBN.

In the next, we follow the Ref.~\cite{NBRResilient} that proved the monotonicity of RBN under local monitoring (unital local map) operation and take a step further by analyzing the RBN behavior under unital and non-unital local noise channels. We use the bipartite Werner state to illustrate that the quantifier $N_{\mathrm{br}}$ is monotonous under general local disturbances. 
Consider the following state
\begin{equation}
\rho_{\mu} = (1-\mu)\frac{\mathbbm{1}^{\mathcal{A}} \otimes \mathbbm{1}^{\mathcal{B}}}{4} + \mu|s \rangle \langle s|,
\label{eq10}
\end{equation}
which is a mixture that interpolates between the singlet pure state $|s \rangle = \frac{|0\rangle |1\rangle - |1\rangle|0\rangle}{\sqrt{2}}$ and a maximally mixed state varying the parameter $\mu$ $\in$ $[0,1]$. We analyzed the action of quantum noise channel acting locally in one of the state partitions, using the  the bit inversion (IB), phase inversion (IF), bit and phase inversion (IBF), depolarization (DP) and amplitude damping (AD) \cite{Chuang}. These channels are mathematically represented in Table 1, where $\mathbbm{1}$ is the identity operator, $\sigma_{1,2,3}$ are the Pauli matrices and, in the case of AD channels, 
$\delta_{0} = $ $\begin{pmatrix}
                   1 & 0 \\
                   0 & s
                 \end{pmatrix}$ and $\delta_{1} = $ $\begin{pmatrix}
                   1 & t \\
                   0 & 0, 
                 \end{pmatrix}$, where $s = \sqrt{1-\gamma}$ and $t = \sqrt{\gamma}$, is the parameters $(p,\gamma)$ $\in$ $[0,1]$ 
the probabilities of noise effects. Moreover, all the inversion channels and the DP channel are unital $\zeta(\mathbbm{1}) = \mathbbm{1}$ while the AD channel is not, that is, $\zeta(\mathbbm{1}) \neq \mathbbm{1}$ \cite{SumeetKunal}.

\begin{table}[!t]
\centering
\setlength{\tabcolsep}{5pt} 
\begin{tabular}{c|cc}\hline \hline
Channels & \multicolumn{2}{c}{Elements of quantum operations} \\ \hline 
IB & $\zeta_{0}=\sqrt{p}\mathbbm{1}$ & $\zeta_{1}=\sqrt{1-p}\sigma_{1}$  \\ 
IF & $\zeta_{0}=\sqrt{p}\mathbbm{1}$ & $\zeta_{1}=\sqrt{1-p}\sigma_{3}$  \\ 
IBF & $\zeta_{0}=\sqrt{p}\mathbbm{1}$ & $\zeta_{1}=\sqrt{1-p}\sigma_{2}$  \\ 
DP & $\zeta_{0}=\sqrt{1-3p/4}\mathbbm{1}$ & $\zeta_{1,2,3}=\sqrt{p/4}\sigma_{1,2,3}$  \\ 
AD & $\zeta_{0,1}=\sqrt{p} \ \delta_{0,1}$ & $\zeta_{2,3}=\sqrt{1-p} \ \delta_{2,3}$ \\ \hline \hline
\end{tabular} 
\caption{Noise channels used in this work. We analyzed the action of the bit inversion (IB), phase inversion (IF), bit and phase inversion (IBF), depolarization channels (DP) and amplitude damping (AD) on the RBN quantified by Eq. \ref{eq5}. Here, the terms $\zeta_{0,1}$ are the elements of quantum operation, where $\mathbbm{1}$ is the identity operator, $\sigma_{1,2,3}$ are Pauli's matrices and $\delta_{0,1}$ are matrices that represent, exclusively, the action of the AD channel. The probabilities of noise occurrence are given by the parameters $(p,\gamma)$ $\in$ $[0,1]$.}
\label{table1}
\end{table}
Using the adopted formalism, we can represent the action of the noise channels on the state of Werner through the map
\begin{equation}
\zeta(\rho_{\mu}) = \sum_{i}(\mathbbm{1}^{\mathcal{A}} \otimes \zeta_{i}^{\mathcal{B}}) \rho_{\mu} (\mathbbm{1}^{\mathcal{A}} \otimes \zeta_{i}^{\mathcal{B}}),
\label{eq11}
\end{equation}
where the noise acts in $\mathcal{B}$. The elements $\zeta_{i}^{\mathcal{A(B)}}$ 
are the terms that model the type of noise that acts on the state $\rho_{\mu}$. 
We observe that in this case, the projective measurement maps $\Phi_{A(B)}$ will act on the noisy state $\zeta(\rho_{\mu})$ to evaluate the RBN. The results presented compare the RBN calculated for Eq. \ref{eq11} and for the state $\rho_{\mu}$ without noise, Eq. \ref{eq10}. This was done with a statistical analysis by randomly selecting several states through a random number generator developed in the \textit{Mathematica 7} software, so that each point of the obtained distribution was calculated for random values of $(\mu, p,\gamma) \in [0,1]$.
\begin{figure}[!t]
 \centering
 \includegraphics[scale=0.15]{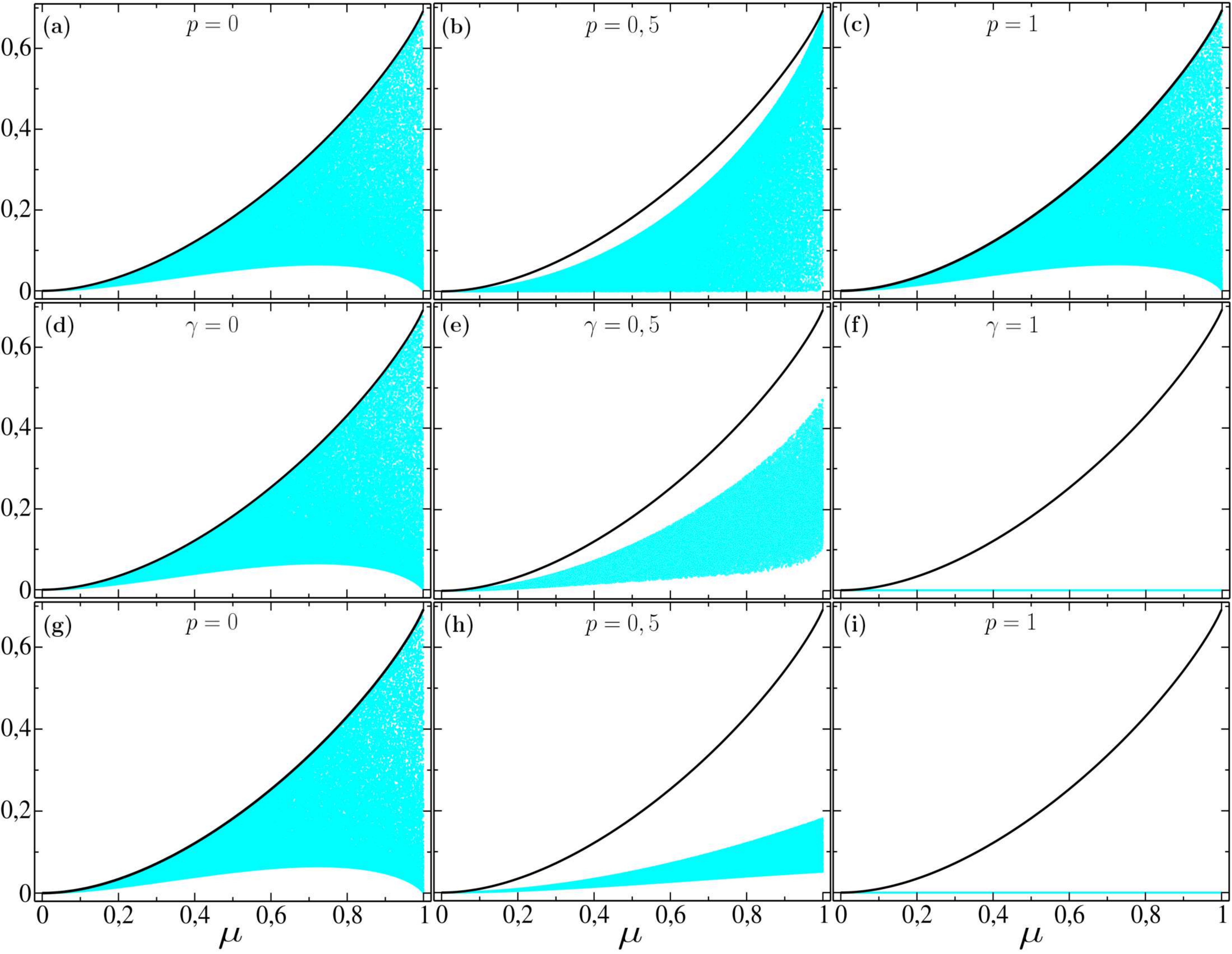}
 \caption{{\footnotesize Illustration of the action of noise channels on the RBN calculated for the state $\rho_{\mu}$. In the first line (a, b, c) we have the representation of the IB channel effect. In the second line, we present the effect of the AD channel. In the third line, the effect of the DP channel. In all the frames, the black curve represents the RBN computed analytically through its definition, Eq. \ref{eq5}, without the action of noise channels. For all the tables above, the numerical calculation was performed for $ 10^{5}$ points, which form the region in cyan, with $p$ and $\gamma$ being the probabilities of the noise occurrence. Each point of the obtained distribution was calculated for random values of $(\mu,p,\gamma)\in[0,1]$.}}
 \label{Fig1}
\end{figure}
In Fig. \ref{Fig1}, the first line shows the effect produced by the inversion channels, IB, IF, and IBF. In this case, both channels produced the same effect due to the rotational invariance of the singlet state, so that in Fig. \ref{Fig1} only the IB channel action is represented. In Fig. \ref{Fig1} (a), we see that in the absence of noise, $p=0$, the numerical result (the region in cyan) coincides with the analytical result (black curve) of the reference \cite{NBR} and is computed here via Eq . \ref{eq5}. 
When adding the effect of the inversion channels, we observed degradation in the RBN as a result of the action of the noise introduced in the system, which is seen in Fig. \ref{Fig1} (b) for $p = 0.5$. As for $p = 1$, we notice that there is no degradation, since the term $ \sqrt{1-p} $ that multiplies the matrices $\sigma_{1,2,3}$ becomes null in the three cases, which eliminates the effect of the inversion channels, Fig. \ref{Fig1} (c).

Consider the effect of the AD channel, seen in the second line of Fig. \ref{Fig1}, we see that for $p=0$ the same discussion as above applies. However, we note that for $p=0.5$, the noise performance causes a more considerable RBN degradation than the effect caused by the inversion channels and, unlike these, the degradation is total for $p=1$, as can be seen in Figs. 1 (e, f). In the third line of Fig. \ref{Fig1}, we can see the case of the DP channel. Note that in relation to the previous cases, the RBN suffers a much more accentuated degradation to $p = 0.5$, Fig. \ref{Fig1} (h), being totally degraded also to $p=1 $, Fig. \ref{Fig1} (i), indicating that the RBN is less resistant to the action of this noise channel.
It is important to note that in all the cases described above, the RBN measurement satisfies the monotonicity condition, as
\begin{equation}
N_{\mathrm{rb}}(\rho_{\mu}) \geqslant N_{\mathrm{rb}}(\zeta(\rho_{\mu})),
\label{eq12}
\end{equation}
showing that the measure provided by the quantifier $N_{\mathrm{rb}}$ is monotonous under local operations as expected from a measure that aims to quantify the nonlocal resources of a quantum state.

\section{RBN as a security witness}

In this section, we argue that is possible to implement RBN as a security witness by demonstrating that this measure is sensitive to external interventions. We compare the scenarios when only the legitimate users of the protocol access the communication channel and when we have the presence of an external agent. 
Defining the legitimate parties of the protocol as Alice and Bob, the procedure can be realized by assuming that Eve (an external agent), will intercepts and measures an observable $B$ in the qubit directed to Bob. Note that, after Eve's intervention the state of the system is described by the map
\begin{equation}
\xi(\rho) = \sum_{b} (\mathbbm{1}^{\mathcal{A}} \otimes B_{b}) \, \rho \, (\mathbbm{1}^{\mathcal{A}} \otimes B_{b}),
\label{eq13}
\end{equation}
with $B = \sum_{b}b\,B_{b}$  an observable acting in $\mathcal{H}_{\mathcal{B}}$ with projectors $B_{b} = |b\rangle \langle b|$ 
described in terms of their eigenstates. The scenario described above provides the following result:

 \textit{For any external intervention on a quantum channel established by a state $\rho$ $\in$ $\mathcal{H}_{\mathcal{A}} \otimes \mathcal{H}_{\mathcal{B}}$ is verify that}
\begin{equation}
0 \leqslant N_{\mathrm{rb}}(\xi(\rho)) \leqslant N_{\mathrm{rb}}(\rho),
\label{eq14}
\end{equation}
\textit{where the equality occurs for totally unrelated states $\rho = \rho_{\mathcal{A}} \otimes \rho_{\mathcal{B}}$}.

The inequality $N_{\mathrm{rb}}(\xi(\rho)) \leqslant N_{\mathrm{rb}}(\rho)$ can be analyzed considering the cases in which maximally incompatible observables are involved in the measurements. In this case, we have observables $\left\lbrace B,B' \right\rbrace$ $\in$ $\mathcal{H}_{\mathcal{B}}$ in which the basis $\left\lbrace |b\rangle \right\rbrace$ and $\left\lbrace |b' \rangle \right\rbrace$ are mutually unbiased so that $\langle b\,|\,b' \rangle = \frac{\mathbbm{1}}{\sqrt{d_{\mathcal{B}}}}$. Thus, the Eq. \ref{eq13} reads
\begin{equation}
\Phi_{B^{'}}(\rho) = \sum_{b^{'}} (\mathbbm{1}^{\mathcal{A}} \otimes B^{'}_{b^{'}}) \, \rho \, (\mathbbm{1}^{\mathcal{A}} \otimes B^{'}_{b^{'}}),
\end{equation}
so that the contextual nonlocality, Eq. \ref{eq4}, is  $\eta_{AB} (\Phi_{B^{'}}(\rho)) = S(\Phi_{A}\Phi_{B^{'}}(\rho)) + S(\Phi_{B}\Phi_{B^{'}}(\rho)) - S(\Phi_{A}\Phi_{B}\Phi_{B^{'}}(\rho)) - S(\Phi_{B^{'}}(\rho))$, where $\Phi_{B}\Phi_{B^{'}}(\rho)$, results in
\begin{eqnarray}
\Phi_{B}\Phi_{B^{'}}(\rho) &=& \sum_{b}(\mathbbm{1}^{\mathcal{A}} \otimes B_{b})\Phi_{B^{'}}(\rho) (\mathbbm{1}^{\mathcal{A}} \otimes B_{b}) \nonumber \\ &=& \frac{\mathbbm{1}}{d_{\mathcal{B}}} \sum_{b^{'}}\langle b^{'}|\rho| b^{'} \rangle \otimes \sum_{b} |b\rangle \langle b| \nonumber \\ &=& \mathrm{Tr}_{\mathcal{B}}\,\rho \otimes \frac{\mathbbm{1}}{d_{\mathcal{B}}} = \rho_{\mathcal{A}} \otimes \frac{\mathbbm{1}}{d_{\mathcal{B}}},
\end{eqnarray}
which also provides
\begin{equation}
\Phi_{A}\Phi_{B}\Phi_{B^{'}}(\rho) = \Phi_{A}(\rho_{\mathcal{A}}) \otimes \frac{\mathbbm{1}}{d_{\mathcal{B}}}.
\end{equation}

 In this way, $\eta_{AB} (\Phi_{B^{'}}(\rho))$ can be rewritten as $\eta_{AB} (\Phi_{B^{'}}(\rho)) = S(\Phi_{A}\Phi_{B^{'}}(\rho)) - \left[ S(\Phi_{A}(\rho_{\mathcal{A}})) - S(\rho_{\mathcal{A}}) \right] - S(\Phi_{B^{'}}(\rho))$, where direct comparison with Eq. \ref{eq4} makes clear that $\xi(\rho) = \Phi_{B^{'}}(\rho)$ implies $\eta_{AB}(\xi(\rho)) \leqslant \eta_{AB}(\rho)$. Furthermore, an extension of $\eta_{AB} (\Phi_{B^{'}}(\rho))$ also shows that
\begin{eqnarray}
\eta_{AB} (\Phi_{B^{'}}(\rho)) &=& S(\Phi_{A}\Phi_{B^{'}}(\rho)) - \Im(A|\rho_{\mathcal{A}})- S(\Phi_{B^{'}}(\rho)) \nonumber \\ &=& \Im(A|\Phi_{B^{'}}(\rho)) - \Im(A|\rho_{\mathcal{A}}),
\label{eq18}
\end{eqnarray}
with $\Im(A|\rho_{\mathcal{A}})$ being a measure of local irreality of $A$ (see ~\cite{Bil} for more details), which is null for any pure entangled state $(|\phi \rangle = \frac{1}{\sqrt{d}} \sum_{i}| i \rangle | i \rangle)$, since the reduced state $\rho_{\mathcal{A}} = \mathrm{Tr}_{\mathcal{B}}\rho = \frac{\mathbbm{1}}{d_{\mathcal{A}}}$ is always maximally mixed with $d_{\mathcal{A}}$ being the dimension of $\mathcal{H}_{\mathcal{A}}$. This clearly shows that $\eta_{AB}(\xi(\rho)) \leqslant \eta_{AB}(\rho)$, which proves the second part of the result of Eq. \ref{eq14}, that is, $N_{\mathrm{rb}}(\xi(\rho)) \leqslant N_{\mathrm{rb}}(\rho)$. The equality occurs for fully uncorrelated states $\rho = \rho_{\mathcal{A}} \otimes \rho_{\mathcal{B}}$, in which case $N_{\mathrm{rb}}(\xi(\rho)) = N_{\mathrm{rb}}(\rho) = 0$, which completes the prove.                                                 
\begin{figure}[!t]
 \centering
 \includegraphics[scale=0.02]{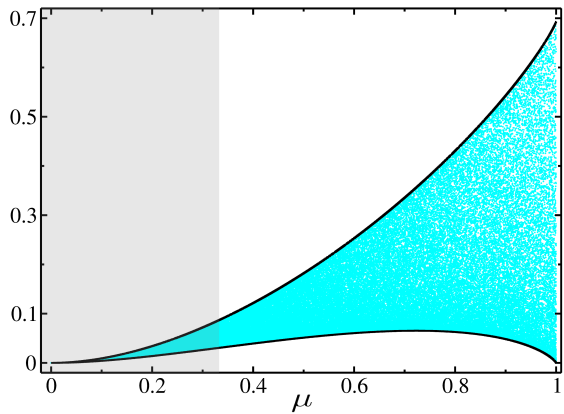}
 \caption{{\footnotesize Scenario that guarantees the security of the quantum channel. The upper black curve is the RBN computed by Alice and Bob when they measure the same observable $\sigma_{j}$. The lower black curve is the lower RBN shared when they measure distinct observables $\sigma_{i}$ and $\sigma_{j}$ com $i,j = 1,2,3$.  The cyan region represents numerical computation of the RBN performed for $10^{5}$ points and considering the parameter range $\mu \, \in \, [0,1]$, the angles $(\theta_{a},\theta_{b}) \in [0,\pi]$, and $(\phi_{a},\phi_{b}) \in [0,2\pi]$, which specify the measurement directions of the legitimate parties. This configuration informs Alice and Bob that the channel in question is safe to perform tasks of transmission and processing of quantum information.}}
 \label{Fig2}
\end{figure}

In the following we apply the previous results by considering that a source emits a bipartite Werner state, Eq. \ref{eq10},  and at each experimental run, the legitimate parties will choose randomly one of three possible observables represented by the Pauli's matrices $\sigma_{i}$ with $i \in (1,2,3)$. We then analyse a situation in which Eve, when intercepting Bob's qubit, measures an observable $B^{'} = \hat{b}^{'} \cdot \vec{\sigma} = \sum_{j}b^{'}_{j}\sigma_{j}$, where $\hat{b^{'}} = \hat{b^{'}}(\theta_{b^{'}},\phi_{b^{'}})$ is the unit vector that specifies the measurement direction of Eve and being $\vec{\sigma} = (\sigma_1,\sigma_2,\sigma_3)$. In contrast, Alice and Bob follow the established protocol and perform only measurements of the three spin observables above. Following what is determined by Eq. \ref{eq18}, the eigenvalues can be computed $\left\lbrace \frac{1-\mu r}{4},\frac{1+\mu r}{4} \right\rbrace$ 
doubly degenerate to $\Phi_{A}\Phi_{B^{'}}(\rho_{\mu})$ and $\left\lbrace \frac{1-\mu}{4},\frac{1+\mu}{4} \right\rbrace$ for $\Phi_{B^{'}}(\rho_{\mu})$. On the other hand, being $\rho_{\mu}^{\mathcal{A}}$ the reduced state of $\mathcal{A}$, we have that $\rho_{\mu}^{\mathcal{A}} = \mathrm{Tr}_{\mathcal{B}}\,\rho_{\mu} = \frac{\mathbbm{1}}{2}$, in a way that $\Phi_{A}(\rho_{\mu}^{\mathcal{A}}) = \frac{\mathbbm{1}}{2}$, which gives
\begin{equation}
\eta_{AB} (\xi(\rho_{\mu})) = \frac{1}{2}\left[ F(\mu) + F(-\mu) - G(\mu,r) - G(-\mu,r) \right], \nonumber
\end{equation}
with $F(\mu) := (1+\mu)\ln\frac{1+\mu}{4}$ and $G(\mu,r) := (1+\mu r)\ln\frac{1+\mu r}{4}$, so that the maximization required for computing the RBN, Eq. \ref{eq5}, is reached for $r=0$, with $r(\theta_{b^{'}},\phi_{b^{'}}) \in [-1,1]$ an angular function that contains information about Eve's measurement directions. Then
\begin{equation}
N_{\mathrm{rb}}(\xi(\rho_{\mu})) = \frac{1}{2}\left[ F(\mu) + F(-\mu) \right] +\ln4,
\label{eq19}
\end{equation}
this result is verified whenever Eve measures any observable that is maximally incompatible with the observables measured by the legitimate parties. In addition, as will be discussed in the sequence, Eq. \ref{eq19} provides the maximum value of $N_{\mathrm{rb}}(\xi(\rho_{\mu}))$, such value always being less than $N_{\mathrm{rb}}(\rho_{\mu})$, as predicted by the result of Eq. \ref{eq14}.

\begin{figure}[!t]
 \centering
 \includegraphics[scale=0.02]{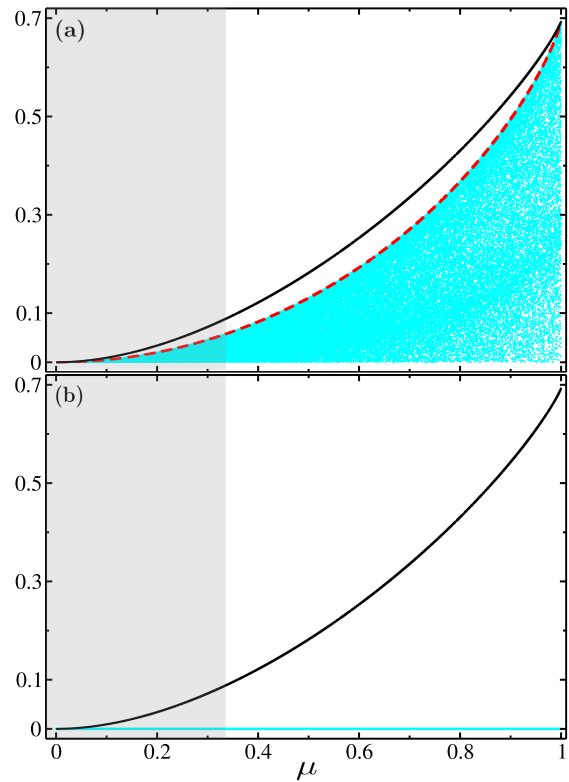}
 \caption{{\footnotesize Scenarios that denounce Eve's interference in the protocol. In Fig. \ref{Fig3}a, we see a complete disfigurement of what Alice and Bob expected to observe after the protocol was performed. The upper black curve is the RBN for the state $\rho_{\mu}$, a result that should be observed when the parts measure the same observable. In the case of distinct observables, we note that the lower share $N_{\mathrm{min}}(\rho_{\mu})$ (discussed in Fig. \ref{Fig2}) is completely degraded due to Eve's interference. In this same configuration, interestingly, we see that degradation of the RBN is verified (red dashed line) when maximally incompatible observables are involved in the measurements of the parties involved. In Fig. \ref{Fig3}b, we observe the total degradation of the RBN verified when the three parts measure the same observable and also when Alice and Bob measure different observables while Eve measures the same observable as one of them.}}
\label{Fig3}
\end{figure}

Another situation is in cases where the parties involved measure the same observables. In these cases Eq. \ref{eq5} points out that $N_{\mathrm{rb}}(\xi(\rho_{\mu})) = 0$, as $\Phi_{\sigma_{j}}\Phi_{\sigma_{j}}(\rho_{\mu}) = \Phi_{\sigma_{j}}(\rho_{\mu})$ for $j=1,2,3$. All the scenarios involved are described in the illustrations in Figs. \ref{Fig2} and \ref{Fig3}, where a statistical analysis performed for the ideal case $\left[N_{\mathrm{rb}}(\rho_{\mu})\right]$  is illustrated in Fig. \ref{Fig2} while the case involving Eve's performance is described in Fig. \ref{Fig3}. Analysis was carried out by randomly selecting several states through a random number generator developed in the software \textit{Mathematica 7}, so that each point of the distribution was calculated for random values of $\mu \in [0,1]$ and the angular variables $\theta_{a(b)} \in [0,\pi]$ and $\phi_{a(b)} \in [0,2\pi]$, which describe the directions of Alice and Bob's measurements, case of Fig. \ref{Fig2}, and $\theta_{e} \in [0,\pi]$ and $\phi_{e} \in [0,2\pi]$ for Eve's measurement directions, Fig. \ref{Fig3}a and Fig. \ref{Fig3}b.

Fig. \ref{Fig2} shows the illustration of the ideal case, a situation that guarantees the security of the quantum channel used in the execution of the protocol. The upper black curve represents the RBN associated with Werner's state, quantified via $N_{\mathrm{rb}}(\rho_{\mu})$, being verified in cases where the parts measure the same observable $\sigma_j$. On the other hand, the lower black curve represents the lower share $N_{\mathrm{min}}(\rho_{\mu})$, obtained when Alice and Bob measure different observables $\sigma_{i}$ and $\sigma_j$ to $i,j = 1,2,3$. The cyan region represents the RBN computed numerically for $\rho_{\mu}$, limited by the upper and lower curves described above. The gray region represents the class of separable states, this is, states that have no entanglement and Bell's nonlocality, but the presence of RBN is notorious in this region.

An analysis of the scenario involving Eve's attacks, Fig. \ref{Fig3}a, clearly shows a degradation of the RBN associated with $\rho_{\mu}$ and a total disfigurement of the picture that Alice and Bob expect to observe. This is observed both by the numerical result, the region in cyan, and by the analytical result described by the red dashed curve. Both results represent situations in which maximally incompatible observables are involved in the measurements of both parties. So, when Eve interferes in the protocol, Alice and Bob do not observe $N_{\mathrm{rb}}(\rho_{\mu})$ when they measure the same observable, and $N_{\mathrm{min}}(\rho_{\mu})$ when they measure different observables, starting to observe a degradation of the RBN associated with $\rho_{\mu}$, contrary to the expected scenario. In addition, it is noted that the lower share is completely degraded due to Eve's interference.

 Finally, Fig. \ref{Fig3}b illustrates the situation mentioned above in which all parts measure the same observable and also situations in which Alice and Bob measure different observables while Eve measures the same observable as one of the two. In both cases, the entire RBN associated with the state is degraded by Eve's attacks, showing that the channel's security has been corrupted. These results corroborate the result stated by Eq. \ref{eq14}, according to which $N_{\mathrm{rb}}(\xi(\rho))$ is non-negative and has a maximum value less than $N_{\mathrm{rb}}(\rho)$, being obtained for observables maximally incompatible. Thus, as shown above, it is possible to use RBN to guarantee the security of a quantum channel that will be used in tasks of quantum communication. A point to be highlighted is the class of separable states included in the interval $0 \leqslant \mu \leqslant 1/3$, represented in the gray region of Figs. \ref{Fig1}, \ref{Fig2} and \ref{Fig3}. It is observed that in this interval, we have the presence of RBN for states that have zero resources such as entanglement and Bell's non-locality.
 
 In what follows, we continue this line of investigation by showing that the results obtained so far in this work, that is, the monotonicity and the hierarchy referring to the RBN, can also bring new perspectives to the use of thermal correlated states for security protocols.

\section{Monotonicity and asymptotic behaviour of RBN for thermal correlated states}

In the section, we investigate from the point of view of thermal correlated states how the RBN is affected by the initial local temperature of each subsystem. The analysis is done after performing an optimal protocol that maximizes the entanglement (concurrence, the relative entropy, and the negativity of
entanglement~\cite{X_State}) on the global thermal system. In contrast to the entanglement which vanishes completely at a finite critical temperature for that state, the global quantum discord (GD) and local quantum uncertainty was shown to decay asymptotically with the temperature~\cite{X_State}. Here, we analyze the RBN and global quantum discord observing that both respects the hierarchic relation \cite{NBR, NBRResilient} of quantum correlation measures and vanished asymptotically for large $T$. We also investigate the effects of quantum noise on the RBN of the considered thermal state and its possible application in secure quantum communication.

Consider  two uncorrelated qubits each one in initially thermal equilibrium state  with the same thermal reservoir at temperature $T$, and the following local Hamiltonian $\sum_{k}E_k\ket{k}\bra{k}$ described in respect with the energy eingenbasis. The corresponding total state is $\rho_0=\tau_{\beta}\otimes\tau_{\beta}$
\begin{equation}
    \tau_{\beta}=\frac{1}{\cal{Z}(\beta)}e^{-\beta H},
\end{equation}
 where $\tau_{\beta}$ is a local  Gibbs state with $\cal{Z}(\beta)$ and $\beta$ being respectively the canonical partition function and the inverse temperature. Consider for simplicity that the
energy of the ground and excited state are respectively $E_0=0$ and $E_1=E$, in such way that the population of the ground state reads
\begin{equation}
    q=\frac{1}{1+e^{-\beta E}},
\end{equation}
and thus the excited state probability is $1-q$, and $q\geq 1/2$ that is the constraint to avoid reservoir with negative temperature. With that, it is clear to see that $\tau_{\beta}=q\ket{0}\bra{0}+(1-q)\ket{1}\bra{1}$, and  the internal energy of some state $\rho$ can be found as $ \expval{U}=\text{Tr}(H \rho)$.
 As discussed in Ref. \cite{X_State}, the thermal correlation can be realized in terms of the unitary evolution $U=V_2V_1$ such that
\begin{equation}
    V_1=\ket{00}\bra{00}+\ket{01}\bra{01}+\ket{11}\bra{10}+\ket{10}\bra{11},
\end{equation}
and
\begin{equation}
    V_2=\ket{\phi_{+}}\bra{00}+\ket{01}\bra{01}+\ket{10}\bra{10}+\ket{\phi_{-}}\bra{11},
\end{equation}
where $\ket{\phi_{+}}=\frac{\ket{00}+\ket{11}}{\sqrt{2}}$ and $\ket{\phi_{-}}=\frac{\ket{00}-\ket{11}}{\sqrt{2}}$.
This results in the following correlated state $\rho_X=U\rho_0U^\dagger$ represented here in the basis $\left\{\ket{00},\ket{01},\ket{10},\ket{11}\right\}$
\begin{equation}
    \rho_X=\left(
\begin{array}{cccc}
 q/2 & 0 & 0 & q^2-q/2 \\
 0 & q(1-q) & 0 & 0 \\
 0 & 0 & (1-q)^2 & 0 \\
  q^2-q/2 & 0 & 0 & q/2 \\
\end{array}
\right),
\end{equation}

In the following we apply the same methods developed in this work to investigate this scenario from the perspective of the  quantifier of RBN and we also use that to highlight the hierarchy of quantum correlations with the Global Discord (GD). This can be done by considering the eigenvectors
\begin{equation}
|\varphi_{+}^{k} \rangle = \cos(\theta_{k}/ 2) |+ \rangle + e^{i\phi_{k}}\sin(\theta_{k}/ 2)|- \rangle
\end{equation}
\begin{equation}
|\varphi_{-}^{k} \rangle = -\sin(\theta_{k}/ 2) |+ \rangle + e^{i\phi_{k}}\cos(\theta_{k}/ 2)|- \rangle,
\end{equation}
where $k = \left\lbrace a, b \right\rbrace$, $\theta_{k} \in [0,\pi]$ e $\phi_{k} \in [0,2\pi]$. 
We then have the measurement projectors $A(B)_{\pm} = |\varphi_{\pm}^{k} \rangle \langle \varphi_{\pm}^{k}|$, and the post-measured states $\Phi_{A}(\rho_{X}) = \sum_{i}(A_{i}\otimes \mathbbm{1}_{\mathcal{B}}) \, \rho_{X} \, (A_{i}\otimes \mathbbm{1}_{\mathcal{B}})$ and $\Phi_{B}(\rho_{X}) = \sum_{i}(\mathbbm{1}_{\mathcal{A}}\otimes B_{i}) \, \rho_{X} \, (\mathbbm{1}_{\mathcal{A}}\otimes B_{i})$, which allows one to evaluate all possible directions in order to find those that provide the RBN and the GD associated with the state $\rho_{X}$.
\begin{figure}[!t]
 \centering
 \includegraphics[scale=0.32]{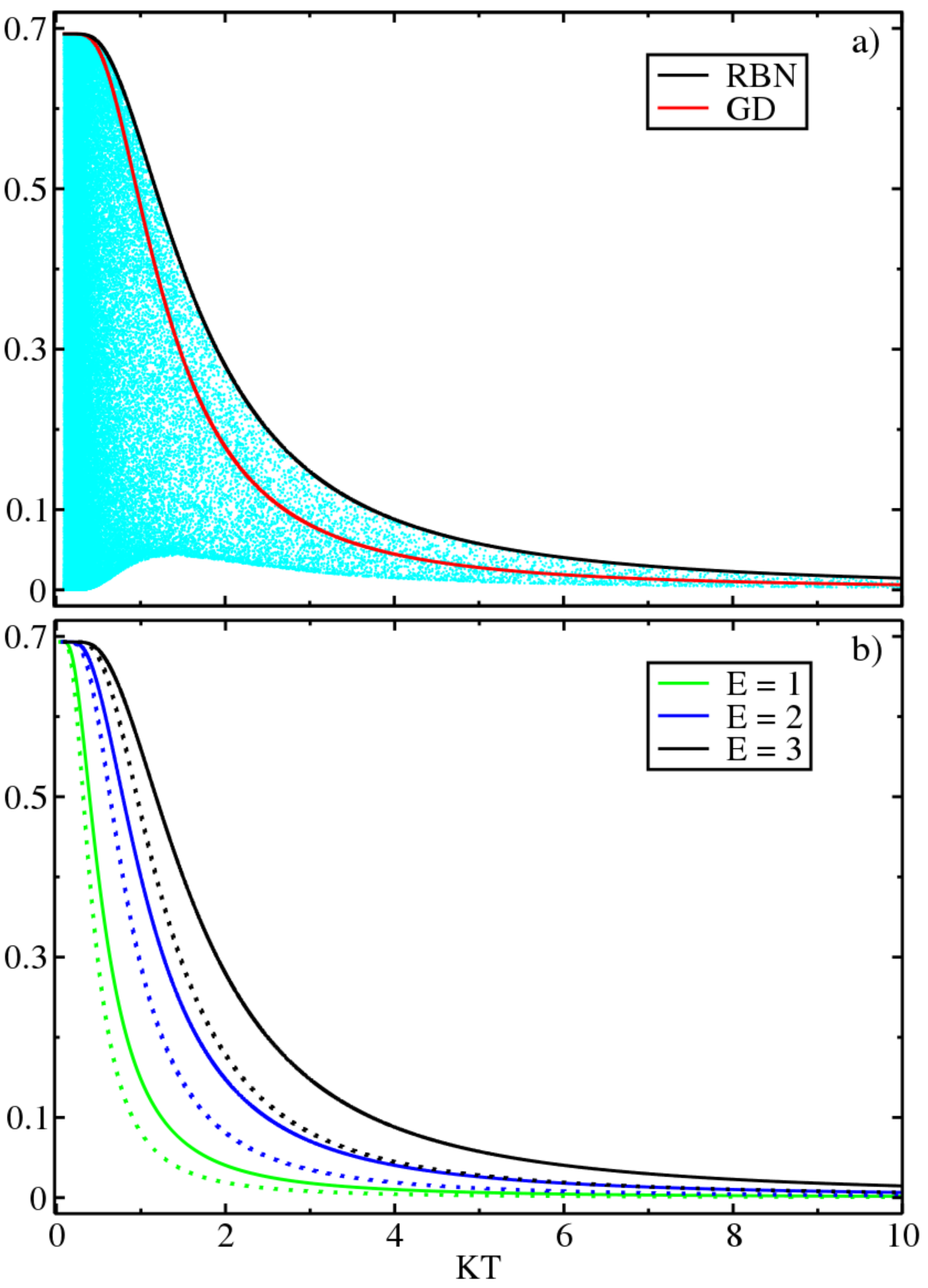}
 \caption{\footnotesize Thermal effect in terms of $KT$ and hierarchy relation between RBN and GD. In Fig. \ref{Fig4}a the cyan region obtained for $E = 3$ correspond to the RBN computed to $\rho_{X}$ state computed numerically for $10^{5}$ points. This region is limited by the computed analytically RBN which is represented by the black curve. The red curve correspond to the analytical global discord $D_{g}$ and also to the context realism-based nonlocality calculated for $A_{\pm} = B_{\pm} = |\varphi_{\pm}^{k} \rangle \langle \varphi_{\pm}^{k}| = \sigma_{z}$. In Fig. \ref{Fig4}b, we show analytically the RBN and GD for $E = 1, 2, 3$. The solid lines are the RBN while dotted lines correspond to GD. Note that as E increases, the quantum correlations contained in $\rho_{X}$ grows. Furthermore, we also note that the hierarchical relationship between these quantitative measures is maintained even at the asymptotic limit where both RBN and GD tend to disappear.}
 \label{Fig4}
\end{figure}

\begin{figure}[!t]
 \centering
 \includegraphics[scale=0.32]{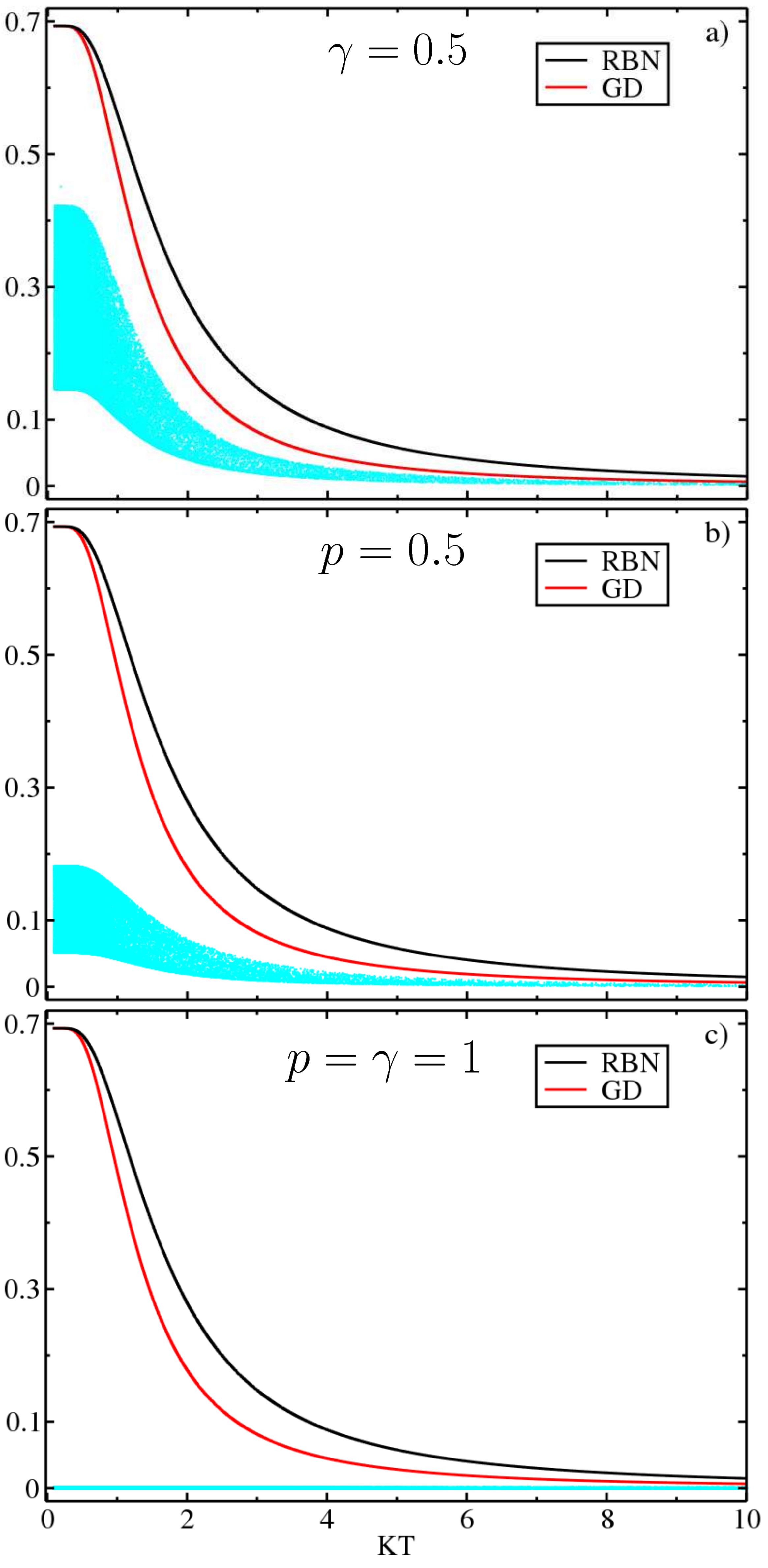}
 \caption{{\footnotesize Effect of the noise channels on the RBN calculated for the thermal state $\rho_{X}$. In Figs. 5a and 5b, we present the effects of the AD and DP channels respectively. In Fig. \ref{Fig4}c, we show the RBN fully degraded for maximum noise intensity $p = \gamma = 1$. For all the frames, the black curve represents the RBN computed analytically, without the action of noise channels while the red curve represents the GD. The numerical calculation was performed for $ 10^{5}$ points, which form the regions in cyan, with $p$ and $\gamma$ being the intensity of the applied noise. }}
 \label{Fig5}
\end{figure}

Analytical calculations supported by numerical investigations showed that the maximization required to compute the RBN associated with the state $\rho_{X}$ 
occurs whenever the pair $(\theta_{k}, \varphi_{k})$ leads to the configuration $A_{\pm} = B_{\pm} = |\varphi_{\pm}^{k} \rangle \langle \varphi_{\pm}^{k}| = \sigma_{x}(\sigma_{y})$, where
\begin{equation}
N_{\mathrm{br}}(\rho_{X}) = \eta_{\sigma_{n}\sigma_{m}}(\rho_{X}) \quad \mathrm{with} \quad  n=m \, \in \, \left\lbrace x,y \right\rbrace,
\end{equation}
while $(\theta_{a},\varphi_{a}) \neq (\theta_{b},\varphi_{b})$ 
always leads to the lowest values of $\eta_{\sigma_{n}\sigma_{m}}$. In what follows we provide a numerical analysis of this result in the Fig. \ref{Fig4}a ($E = 3$) 
where the black curve is the analytically computed RBN for the observables $\sigma_{x(y)}$ 
while the cyan region corresponds to the numerical result obtained through a statistical analysis performed for $10^{5}$ points, which was always limited by the analytical result (black curve) of $\mathrm{N_{br}}(\rho_{X})$. The red curve represents $\eta_{\sigma_{z}\sigma_{z}}(\rho_{X})$ obtained when the observable $\sigma_{z}$ is measure in $\mathcal{H}_{\mathcal{A}}$ e $\mathcal{H}_{\mathcal{B}}$. 
This result is completely analogous to global discord (GD) obtained in \cite{X_State} for $\rho_{X}$, indicating that for $\theta_{a} = \theta_{b} = 0$ both $D_{\mathrm{g}}(\rho_{X})$ and $\eta_{\sigma_{z}\sigma_{z}}(\rho_{X})$ capture the same amount of non-local aspects contained in the considered state.

The discussion about the hierarchy among several measures of quanticity was addressed in Refs. \cite{NBR,NBRResilient}, and the reasoning for the construction of the hierarchy relationship between the RBN and the global discord was based on the observation that a classical-classical state $\rho_{cc} = \sum_{i}p_{a,b}A_{a} \otimes B_{b}$ 
for which it turns out $N_{\mathrm{br}}(\rho_{cc}) > 0 $, while this is a state for which it is well known that $D_{\mathrm{g}}(\rho_{cc}) = 0$, 
what takes states with $N_{\mathrm{br}} > 0$ as a strict superset of measures of quanticity. The results obtained in this section corroborate the hierarchy relationship, as can be seen in Fig. \ref{Fig4}b, where we observe the RBN (solid lines) and the global discord (dashed lines) obtained for $E = 1, 2, 3$. Furthermore, we also observe that increasing the temperature $T$ also decreases the amounts of RBN and global discord, with asymptotically decaying for large $T$.

Finally, we evaluate the effect of noise channels acting on $\mathcal{B}$. In Fig. \ref{Fig5}, we consider only the action of the depolarization (DP) and amplitude damping (AD) channels, showing that the RBN tends to decrease as the noise intensity increases until it cancels out completely for $p =\gamma=0$. As in the previous case, this shows that local actions on state partitions do not increase the RBN computed by $N_{\mathrm{br}}$, so it is caused to degrade or remain unchanged as required for a quantifier of quantum correlations. 

\section{Conclusion}

In relation to the same scenarios related to Bell's nonlocality and entanglement quantifications, we performed in this work an analysis that shows the possibility to elect the RBN quantifier as a possible tool to perform quantum communication tasks, since it generally does not increase under local operations. We show analytically that the $N_{\mathrm{rb}}$ quantifier is invariant under local unitary operations, confirming the unitary invariance property, and we explored numerically, using two-qubits Werner and thermal correlated states, that this measure is monotonous under local operations represented by the local actuation of quantum noise. The verification of these properties, together with the properties verified in \cite{NBRResilient}, show that RBN can be a promising alternative to the other known measures of quantum correlations.

 We proposed a situation in which the RBN acts as a security witness in a bipartite quantum channel represented by the Werner state where two legitimate users, Alice and Bob, certify the channel's security before using it to transmit information. Security certification is performed by the RBN computation when Alice and Bob randomly measure one of Pauli's three observables in several experimental runs. Our calculations showed that the security of the channel is guaranteed when users coincidentally measure the same observable, which provides the RBN associated with Werner's state. Furthermore, when measuring different observables, the lowest value for the RBN of the respective state is verified. On the other hand, in cases where an external agent, Eve, secretly acts on one of the subsystems, we found that RBN's computation works as an indicator that the channel's security has been compromised, as the RBN fails to show the expected pattern. Furthermore, as can be seen from Figs. \ref{Fig2} and \ref{Fig3}, RBN is present even in states without any entanglement and Bell non-locality, which makes RBN a promising alternative in quantum communication tasks. 

In this work, we also investigated the behavior of RBN in optimally correlated thermal states. We confirmed a hierarchical relationship between that the quantifier $N_{\mathrm{br}}(\rho_{X})$ and the global quantum discord $D_{\mathrm{g}}(\rho_{X})$, which shows that it is possible in principle to have RBN even for non-discordant states. The RBN shown to have an asymptotically decay with the initial temperature of the local Gibbs state.  Moreover, we extended the results concerning the monotonicity of RBN under local quantum operations to that scenario of thermal correlated states. This is important to highlight that this measure of quantum correlations can also be used to gain new insights for communications tasks involving thermal correlated states.

\section*{Acknowledgments}

 The authors acknowledge Renato Angelo and Ismael Paiva for helpful comments and suggestions. The authors acknowledge the Brazilian funding agencies FUNCAP and CNPq (grant number 305712/2019-5, V.S.G);  and CAPES (grant number 88887.354951/2019-00, P.R.D).


\begin{thebibliography}{00}

\bibitem{EPR} A. Einstein, B. Podolsky, and N. Rosen, Can quantum-mechanical description of physical reality be considered complete?, \href{https://doi.org/10.1103/PhysRev.47.777}{Phys. Rev. {\bf 47}, 777 (1935)}.

\bibitem{Bell}
J. S. Bell. On the Einstein-Podolsky-Rosen paradox, \href{https://doi.org/10.1103/PhysicsPhysiqueFizika.1.195}{Physics 1, \textbf{195}, (1964)}.

\bibitem{BrunnerWehner}
N. Brunner \textit{et all}. Bell nonlocality, \href{https://doi.org/10.1103/RevModPhys.86.419}{Rev. Mod, Phys., \textbf{86}, 839, (2014)}.

\bibitem{hensen15} B. Hensen {\it et al}.,Loophole-free Bell inequality violation using electron spins separated by 1.3 kilometres. \href{https://doi.org/10.1038/nature15759}{Nature (London) {\bf 526}, 682 (2015)}.

\bibitem{giustina15} M. Giustina {\it et al}., Significant-Loophole-Free Test of Bell's Theorem with Entangled Photons, \href{https://doi.org/10.1103/PhysRevLett.115.250401}{Phys. Rev. Lett. {\bf 115}, 250401 (2015)}.

\bibitem{shalm15} L. K. Shalm {\it et al}., Strong Loophole-Free Test of Local Realism, \href{https://doi.org/10.1103/PhysRevLett.115.250402}{Phys. Rev. Lett. {\bf 115}, 250402 (2015)}.

\bibitem{WisemanDoherty}
H. M. Wiseman, S. J. Jones, and A. C. Doherty. Steering, Entanglement, Nonlocality, and the Einstein-Podolsky-Rosen Paradox, \href{https://doi.org/10.1103/PhysRevLett.98.140402}{ Phys. Rev Lett. {\bf 98}, 140402, (2007)}.

\bibitem{CostaAngelo}
A. C. S. Costa, M. W. Beims, R. M. Angelo. Generalized discord, entanglement, Einstein-Podolsky-Rosen steering, and Bell nonlocality in two-qubit systems under (non-)Markovian channels: Hierarchy of quantum resources and chronology of deaths and births, \href{https://doi.org/10.1016/j.physa.2016.05.068}{Phys. A. {\bf 461}, 469-479, (2016)}.




\bibitem{Bil}
A. L. O. Bilobran and R. M. Angelo, A measure of physical reality, \href{https://doi.org/10.1209/0295-5075/112/40005}{Europhys. Lett. {\bf 112}, 40005 (2015)}.


\bibitem{Monitoring} P. R. Dieguez, R.M. Angelo, Information-reality complementarity, The role of measurements and quantum reference frames, \href{https://doi.org/10.1103/PhysRevA.97.022107}{Phys. Rev. A {\bf97}, 022107 (2018)}.



\bibitem{NBRQuantumWalking}
A. C. Orthey Jr. and R. M. Angelo, Nonlocality, quantum correlations, and violations of classical realism in the dynamics of two noninteracting quantum walkers, \href{https://doi.org/10.1103/PhysRevA.100.042110}{Phys. Rev. A {\bf 100},  04110 (2019)}.

\bibitem{costa20}
A. C. S. Costa and R. M. Angelo, Information-based approach towards a unified resource theory, \href{https://doi.org/10.1007/s11128-020-02826-y}{Quantum Info. Process. {\bf 19}, 325  (2020)}.

\bibitem{Lustosa20} F. R. Lustosa, P. R. Dieguez, and I. G. da Paz, Irrealism from fringe visibility in matter-wave double-slit interference with initial contractive states, \href{https://doi.org/10.1103/PhysRevA.102.052205}{Phys. Rev. A. {\bf 102}, 052205 (2020)}.

\bibitem{Dieguez21} P. R. Dieguez, J. R. Guimar\~{a}es, J. P. S. Peterson, R. M. Angelo, and R. M. Serra, Experimental assessment of physical realism in a quantum controlled device, \href{https://arxiv.org/abs/2104.08152}{arXiv:2104.08152, (2021)}.



\bibitem{NBR}
V. S. Gomes, R. M. Angelo, Nonanomalous realism-based measure of nonlocality, \href{https://doi.org/10.1103/PhysRevA.97.012123}{Phys. Rev. A {\bf 97}, 012123 (2018)}.


\bibitem{NBRResilient} V. S. Gomes, R. M. Angelo, Resilience of realism-based nonlocality to local disturbance, \href{https://doi.org/10.1103/PhysRevA.99.012109}{Phys. Rev. A. {\bf 99}, 012109 (2019)}.


\bibitem{dieguez2018} P. R. Dieguez, R. M. Angelo, Weak quantum discord, \href{https://doi.org/10.1007/s11128-018-1963-1}{Quantum Info. Process. {\bf 17}, 194 (2018)}.


\bibitem{NBRTripartite}
D. M. Fucci and R. M. Angelo, Tripartite realism-based quantum nonlocality, \href{https://doi.org/10.1103/PhysRevA.100.062101}{Phys. Rev. A {\bf 100}, 062101 (2019)}.

\bibitem{PopescuRohrlich}
Sandu Popescu and Daniel Rohrlich. Thermodynamics and the measure of entanglement, \href{https://doi.org/10.1103/PhysRevA.56.R3319}{Phys. Rev. A, \textbf{56}, 5, (1997)}.

\bibitem{GuifreVidalTarrach}
Guifr\'{e} Vidal and Rolf Tarrach. Robustness of entanglement, \href{https://doi.org/10.1103/PhysRevA.59.141}{Phys. Rev. A, \textbf{59}, 1, (1999)}.


\bibitem{BennettWootters}
Charles H. Bennett \textit{et all}. Mixed-state entanglement and quantum error correction, \href{https://doi.org/10.1103/PhysRevA.54.3824}{Phys. Rev. A., \textbf{54}, 3824, (1996)}.

\bibitem{MonrasIlluminati}
A. Monras \textit{et all}. Entanglement quantification by local unitary operations, \href{https://doi.org/10.1103/PhysRevA.84.012301}{ Phys. Rev. A., \textbf{84}, 012301, (2011)}.

\bibitem{VedralPlenio}
V. Vedral and M. B. Plenio. Entanglement measures and purification procedures, \href{https://doi.org/10.1103/PhysRevA.57.1619}{Phys. Rev. A,\textbf{57}, 3, (1998)}.

\bibitem{Horodecki}
R. Horodecki \textit{et all}. Quantum entanglement, \href{https://doi.org/10.1103/RevModPhys.81.865}{Rev. Mod. Phys., \textbf{81}, 865, (2009)}.

\bibitem{GuifreVidal}
Guifr\'{e} Vidal. Entanglement monotones, \href{https://doi.org/10.1080/09500340008244048}{J. Mod. Opt., \textbf{47}, 2/3, (2000)}.

\bibitem{BritoAmaralChaves}
S. G. A. Brito and B. Amaral and R. Chaves. Quantifying Bell nonlocality with the trace distance, \href{https://doi.org/10.1103/PhysRevA.97.022111}{J. Phys. A, \textbf{97}, (2018)}.

\bibitem{JulioVicente}
Julio I de Vicente. On nonlocality as a resource theory and nonlocality measures, \href{https://doi.org/10.1088/1751-8113/47/42/424017}{J. Phys. A: Math. Theor., \textbf{47}, 424017, (2014)}.

\bibitem{GallegoAolita}
Rodrigo Gallego and Leandro Aolita. Nonlocality free wirings and the distinguishability between {B}ell boxes, \href{https://doi.org/10.1103/PhysRevA.95.032118}{Phys. Rev. A, \textbf{95}, 424017(18pp), (2017)}.

\bibitem{OllivierZurek}
 H. Ollivier and W. H. Zurek, Quantum discord: A measure of the quantumness of correlations, \href{https://doi.org/10.1103/PhysRevLett.88.017901}{Phys. Rev. Lett. \textbf{88}, 017901 (2001)}.
 
 \bibitem{Henderson01} L. Henderson and V. Vedral, Classical, quantum and total correlations, \href{https://doi.org/10.1088/0305-4470/34/35/315}{J. Phys. A: Math. Gen. \textbf{34}, 6899 (2001)}; V. Vedral, Classical correlations and entanglement in quantum measurements, \href{https://doi.org/10.1103/PhysRevLett.90.050401}{Phys. Rev. Lett. \textbf{90}, 050401 (2003)}.

\bibitem{RulliSarandy}
C. C. Rulli and M. S. Sarandy. Global quantum discord in multipartite systems, \href{https://doi.org/10.1103/PhysRevA.84.042109}{Phys. Rev. A, \textbf{84}, 042109, (2011)}.

\bibitem{Sapienza19} F. Sapienza, F. Cerisola, and A. J. Roncaglia, Correlations as a resource in quantum thermodynamics, \href{https://doi.org/10.1038/s41467-019-10572-8}{Nat Commun {\bf 10}, 2492 (2019)}. 

\bibitem{X_State}
N. Behzadi, E. Soltani and E. Faizi. Thermodynamic Cost of Creating Global Quantum Discord
and Local Quantum Uncertainty, \href{https://doi.org/10.1007/s10773-018-3838-8}{Int. J. of Theo. Phys., \textbf{57}, 3207--3214, (2018)}.



\bibitem{NewtonI} E. Newton et al, Quantum secrecy in thermal states, \href{https://doi.org/10.1088/1361-6455/ab1e91}{J. Phys. B: At. Mol. Opt. Phys. {\bf 52}, 125501 (2019)}.

\bibitem{NewtonII} E. Newton et al, Quantum secrecy in thermal states II, \href{https://doi.org/10.1088/1361-6455/aba7e9}{J. Phys. B: At. Mol. Opt. Phys. {\bf 53} 205502 (2020)}.



\bibitem{AcinEtAl}
A. Ac\'{i}n \textit{et al}. Quantum nonlocality in two three-level systems, \href{https://doi.org/10.1103/PhysRevA.65.052325}{Phys. Rev. A, \textbf{65}, 052325, (2002)}.

\bibitem{AnomalyNonlocality}
A. A. M\'{e}thot and V. Scarani. An anomaly of non-locality, \href{https://doi/10.5555/2011706.2011716}{Quantum Inf. Comput., \textbf{7}, 157, (2007)}.

\bibitem{VolumeViolacao}
E. A. Fonseca and Fernando Parisio. Measure of nonlocality which is maximal for maximally entangled qutrits, \href{https://doi.org/10.1103/PhysRevA.92.030101}{Phys. Rev. A, \textbf{92}, 030101(R), (2015)}.

\bibitem{Sakurai}
J. J. Sakurai and Jim Napolitano. \textit{Modern Quantum Mechanics}. 2ed. New York: Pearsons, 2010.


\bibitem{Chuang}
Michael A. Nielsen and Isaac L. Chuang. \textit{Quantum Computation and Quantum Information}. 7ed. New York: Cambridge University Press, 2010.

\bibitem{SumeetKunal}
Sumeet Khatri, Kunal Sharma and Mark M. Wilde. Information-theoretic aspects of the generalized amplitude-damping channel, \href{https://doi.org/10.1103/PhysRevA.102.012401}{Phys. Rev. A, \textbf{102}, 012401, (2020)}.




\end{thebibliography}
\end{document}